\documentstyle[aps,12pt,epsf]{revtex}

\input amssym.tex

\def\fun#1#2{\lower3.6pt\vbox{\baselineskip0pt\lineskip.9pt
  \ialign{$\mathsurround=0pt#1\hfil##\hfil$\crcr#2\crcr\sim\crcr}}}
\def\apj{{\it Astrophys. J.}}

\def\etal,{{\it et al., }}

\begin{document}
\draft
%\special{papersize=8.5in,11in}

\title{Transformation-Induced Nonthermal Neutrino Spectra and
\\Primordial Nucleosynthesis} 
\author{Kevork Abazajian, Xiangdong Shi, and
George M. Fuller} 
\address{Department of Physics, University of
California, San Diego, La Jolla, California 92093-0350}
\date{September 17, 1999}

\maketitle

\begin{abstract} 
We examine in detail the changes in the production of primordial
helium resulting from nonthermal neutrino momentum distributions
produced by resonant transformation of electron-type neutrinos to
steriles.  These transformations, $\bar{\nu}_e\rightarrow \bar{\nu}_s$
($\nu_e\rightarrow\nu_s$), amplify a positive (negative) lepton number
asymmetry.  We find that the resulting supression relative to a
thermal distribution of low energy $\nu_e$ reduces $n\rightarrow p$
conversion to a greater extent than does the enhancement of
$n\rightarrow p$ from an identical suppression of $\bar{\nu}_e$.
Thus, equal lepton-number asymmetries of opposite sign have unequal
effects on the resulting helium yield in primordial nucleosynthesis.
\end{abstract}
\bigskip

\newpage 

\section{Introduction}

The role of neutrinos in primordial nucleosynthesis has been exploited
to infer constraints on the number of leptons and light neutrinos in
the Standard Model of particle physics \cite{nnu2}.  Complimentary to
this, the width of the $Z_0$ was found to allow only three light
weakly-coupled neutrinos \cite{zdecay}. This experimental
determination of the number of neutrinos, $N_\nu$, has led some to
describe primordial nucleosynthesis as determined by only one
parameter: the baryon-to-photon ratio, $\eta$. Along with the
assumptions of homogeneity and isotropy, the theory of Standard Big
Bang Nucleosynthesis (SBBN) predicts the abundance of the lightest
nuclides within reasonable error through the single parameter $\eta$.

Now, however, indications of neutrino oscillations from the atmospheric muon
neutrino deficit, the solar neutrino deficit and the LSND excess may
prove that our understanding of neutrino physics is incomplete
\cite{caldconrad}.  This uncertainty in neutrino physics threatens to
destroy the aesthetics of a one parameter SBBN with the several
parameters of the neutrino mass and mixing matrix.

Neutrino mixing in the early universe can induce effects including
matter-enhanced neutrino flavor transformation, partial thermalization
of extra degrees of freedom, and lepton number generation.  Efforts
are sometimes made to crowd the multiple effects of neutrino mixing
into a single parameter, $N_\nu^{\rm eff}$, the effective number of
neutrinos.

The kinetic equations describing the effects of neutrino mixing in the
early universe were first discussed by Dolgov \cite{dolgov81}.  Larger
$\delta m^2$ and $\sin^2 2\theta$ accelerate population of sterile
states.  Resonant conversion of active neutrinos to sterile neutrinos
may also take place when the mass eigenvalue corresponding to $\nu_s$
is smaller than that for the $\nu_\alpha$, {\it i.e.}, $m_s^2 -
m_\alpha^2 = \delta m_{s\alpha}^2 < 0$.  Both resonant and
non-resonant conversion has been used to place constraints on
active-sterile neutrino properties \cite{barbdolg,enqvist,xsf}.

Several formulations of neutrino dynamics in the early universe evolve
the system at a monochromatic average neutrino energy with
interactions occuring at an integrated rate.  This approach is
appropriate when the time scale of interactions between the neutrinos
and the thermal environment is much shorter than the dynamic time
scale of neutrino flavor transformations.  However, when
tranformations and the Hubble expansion are more rapid than
thermalization, neutrino energy distributions can be distorted from a
Fermi-Dirac form.  Furthermore, such spectral distortions can
survive until nucleosynthesis \cite{SMF}.  The role of neutrino energy
distributions in primordial nucleosynthesis has been discussed since
the earliest work on the subject \cite{alpher,wfh}.  Recent
work \cite{kirchiz,xfalept} has emphasized the importance of neutrino
energy distributions on the weak nucleon interconversion rates
$(n\rightleftharpoons p)$.  Kirilova and Chizhov \cite{kirchiz}
included the distortion of neutrino spectra from nonresonant
transformations in their nucleosynthesis calculation.  A time-evolving
distortion was also included in a calculation of helium production
after a neutrino-mixing-generated lepton asymmetry in electron-type
neutrinos from $\nu_e\rightleftharpoons\nu_s$ or
$\bar{\nu}_e\rightleftharpoons\bar{\nu}_s$ (active-sterile)
mixing\cite{xfalept}.

In this paper we explore in detail the effects on primordial
nucleosynthesis of the nonthermal neutrino spectra produced by the
resonant transformations discussed in \cite{xfalept}.  The resonant
transformation of electron-type neutrinos to steriles can leave a near
absence of neutrinos in the low energy end of the active neutrino
energy spectrum.  We find that this suppression of low energy
neutrinos or anti-neutrinos in the otherwise Fermi-Dirac spectrum
alters the neutron-to-proton weak interconversion reactions.  These
alterations stem from initial and final phase space effects.

Specifically, we find that the absence of low energy electron
anti-neutrinos lifts the Fermi blocking of final state neutrinos in
neutron decay.  In addition, the absence of low energy electron
neutrinos capturing on neutrons significantly alters the rate of this
interaction.  Both cases lead to an alteration of the
neutron-to-proton ratio at weak freeze out, which is directly related
to the produced mass fraction of primordial helium.

\section{The Dynamics of Neutrino Transformation}

The general equations describing the evolution of $M$ flavors of
neutrinos in a high density environment such as the early universe are
derived from the evolution equation for the density operator
\begin{equation}
\label{hberg}
i {\partial \rho \over \partial t} = [H,\rho].
\end{equation}
Here $\rho$ can be the density operator for the entire system of $M$
neutrinos or any smaller neutrino system that evolves rather
independently of the other neutrino mixings.  Several authors
\cite{dolgov81,barbdolg,kirchiz} explicitly follow the evolution of
neutrinos through the density matrix.

The evolution of the density operator for a two state active-sterile
system has been popularized in a vector formalism where the density
operator is
\begin{equation}
\label{rhoP}
\rho = {1 \over 2} P_0(1 + {{\bf P} \cdot \sigma}) \;\;\;\;\;\;
\bar{\rho} = {1 \over 2} \bar{P}_0(1 + {{\bf \bar{P}} \cdot \sigma}) 
\end{equation}
where $\bar{\rho}$, $\bf \bar{P}$ and $\bar{P}_0$ refer to the
anti-neutrino states.  For an active-sterile two state system, the
Hamiltonian of eqn. (\ref{hberg}) includes forward scattering off of
the $e^\pm$ background, and elastic and inelastic scattering of
neutrinos from all leptons (see, {\it e.g.}, Ref. \cite{mckthom94}).
For this system, substituting (\ref{rhoP}) into the evolution equation
resulting from Eqn. \ref{hberg} gives
\begin{equation}
\label{masterP}
{d \over dt}{\bf P} = \langle{\bf V}\rangle\times {\bf P} +
(1-P_z)\left({d \over dt} \ln P_0\right)\hat{\bf z} - \left(D^E + D^I
+ {d \over dt}
\ln P_0\right)(P_x\hat{\bf x} + P_y\hat{\bf y})
\end{equation}
\begin{equation}
\label{masterP2}
{d \over dt} P_0 = \sum_{\alpha=e,\nu_\mu,\nu_\tau}
\langle \Gamma (\nu_e\bar{\nu}_e\rightarrow
\alpha\bar{\alpha})\rangle (\lambda_\alpha
n_\alpha n_{\bar{\alpha}}-n_{\nu_e}n_{\bar{\nu}_e})
\end{equation}
where an average over the momentum ensemble of neutrinos is taken for
the vector potential $\bf V$ and the interaction rates $\Gamma$. Here
$D^I$ and $D^E$ are the damping coefficients due to inelastic and
elastic neutrino scattering off of $e^\pm$ and themselves (see
\cite{xsf}). Also, $\lambda_\nu = {1\over 4}$ and $\lambda_e = 1$.
This set of equations (\ref{masterP},\ref{masterP2}) can be followed
for specific neutrino momenta.  An initial thermal neutrino spectrum
can be discretized into $\cal N$ energy bins.  Along with the number
density evolution equations of neutrinos that do not undergo
transformation, these equations form $8\cal N$ equations describing
the evolution of the system.

Here we assume that the sterile neutrino sea is unpopulated at a
temperature $T\sim 100\, \rm MeV$ (see Ref. \cite{fullmalan}). Active
neutrinos can undergo resonant transformation to sterile neutrinos as
the density of $e^\pm$ decreases with the expansion.  As in the sun,
the resonance point is energy dependent. For example, the resonance,
for $\delta m_{s\alpha}^2 < 0$, occurs at
\begin{equation}
\left({E\over T}\right)_{\rm res} \approx {|\delta m^2/{\rm eV^2}|\over
16(T/{\rm Mev})^4 L_{\nu_\alpha}},
\end{equation}
where $E/T$ is the neutrino energy scaled by the ambient temperature.
This resonance adiabatically converts $\nu_\alpha(\bar{\nu}_\alpha)$
to $\nu_s(\bar{\nu}_s)$ as long as the timescale of the resonance is
much longer than the neutrino oscillation period at resonance, and the
resonance position must move slowly through the neutrino energy spectrum (see
Ref. \cite{xfalept}).

As the universe expands and the ambient temperature decreases, the
resonance moves from lower to higher neutrino energies.  The
energy-dependence of the resonance becomes important at temperatures
below $T_{\rm dec} \approx 3 \,\rm MeV$ \cite{SMF}.  Below this
temperature, the neutrinos decouple from the $e^\pm$ background.  The
weak interaction rates of elastic and inelastic scattering of
neutrinos above $T_{\rm dec}$ are much faster than the expansion time
scale.  In this case, the neutrino energy distributions can reshuffle
into a thermal Fermi-Dirac spectrum before the onset of
nucleosynthesis.  However, if the transformation occurs after
decoupling, the expansion rate will be too rapid, and therefore the
neutrino spectra enter the epoch of nucleosynthesis in a non-thermal
state.  This will be the case, usually, when the mass-squared
difference between mass eigenvalues corresponding to $\nu_e$ and
$\nu_s$ is $|\delta m^2_{es}| \leq 1\,\rm eV^2$.

Direct $\nu_e\rightarrow \nu_s(\bar{\nu}_e\rightarrow \bar{\nu}_s)$
transformation was explored by Foot and Volkas \cite{foot} as a
possible means to allow for a sterile neutrino within SBBN with minimal
modification to the theory, and also to reduce the difference between
the $^4$He abundance predicted by the observed primordial deuterium
abundance \cite{burtyt} and the primordial $^4$He abundance inferred
by Olive {\it et al.}  \cite{OSS}.  The change in primordial helium
production in this $\nu_e\rightarrow \nu_s(\bar{\nu}_e\rightarrow
\bar{\nu}_s)$ scenario, including alterations of neutrino spectra, was
described in \cite{xfalept}.  The resonant transformation of
$\nu_e\rightarrow \nu_s(\bar{\nu}_e\rightarrow \bar{\nu}_s)$ produces
a lepton number asymmetry in electron-type neutrinos,
$L_{\nu_e}$. Here we define the net lepton number residing in neutrino
species $\nu_\alpha$ to be $L_{\nu_\alpha} \equiv
{(n_{\nu_\alpha}-n_{\bar{\nu}_\alpha})/{ n_\gamma}}$, where
$n_{\nu_\alpha}$, $n_{\bar{\nu}_\alpha}$, and $n_\gamma = (2
\zeta(3)/\pi^2) T^3$ are the proper number densities of
$\nu_\alpha$, $\bar{\nu}_\alpha$, and photons, respectively.  The sign
of this asymmetry depends sensitively on initial thermal conditions
and can be different in causally disconnected regions
\cite{xdchaos,xdcaus}. In the following sections, we address how
non-thermal spectra affect the production of primordial helium.

\section{Weak Nucleon Interconversion Rates and Primordial Helium}

\begin{figure}[ht]      % in second brace, h=here, t=top, b=bottom      
\centerline{\epsfxsize 4 truein
\epsfbox{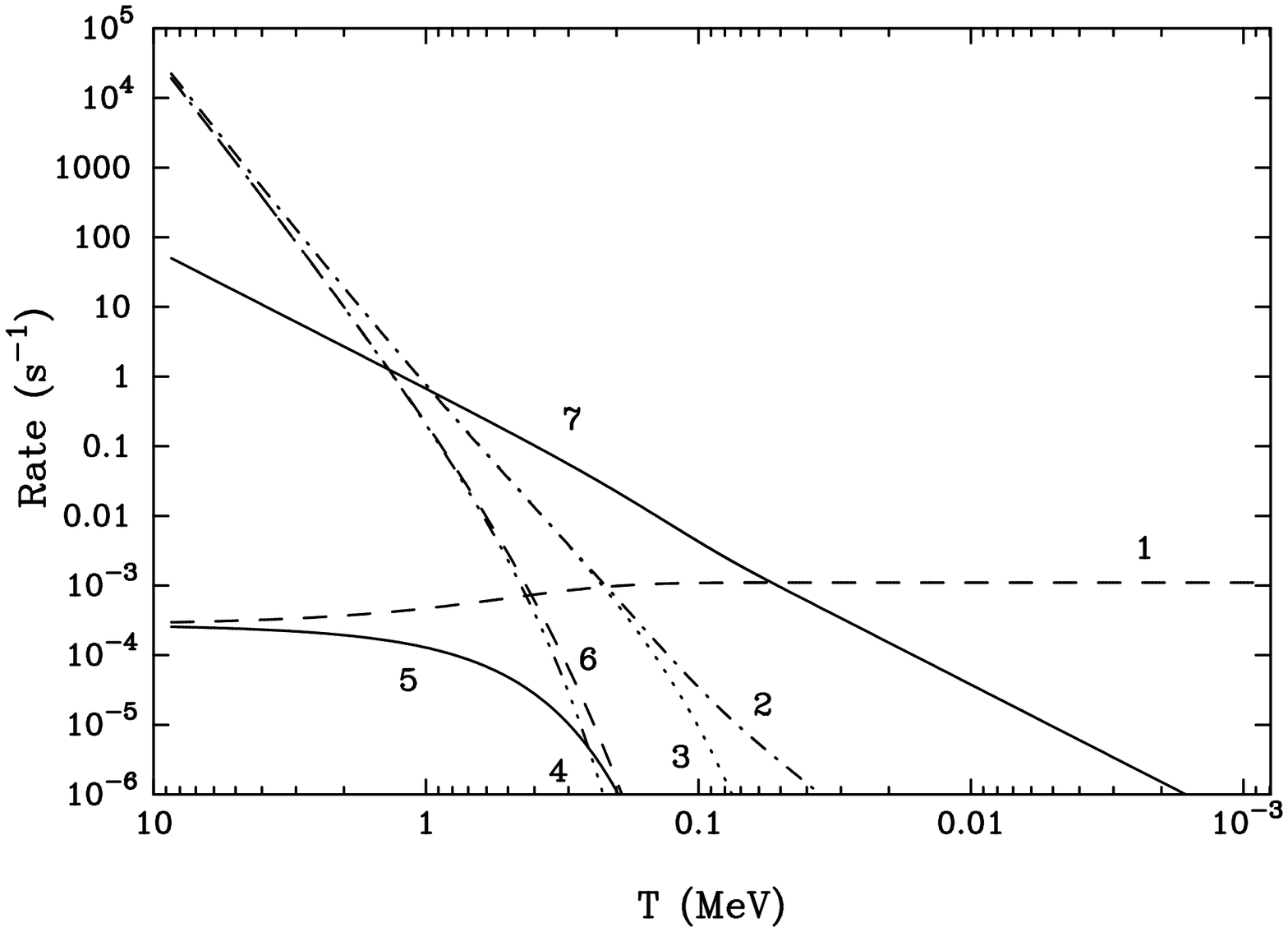}}
\vskip 0 cm
\caption[]{
\small The evolution of the $n\rightleftharpoons p$ rates are shown. 1)
$n \rightarrow p + e^- + \bar{\nu}_e$, 2) $n + \nu_e \rightarrow p +
e^-$, 3) $n + e^+ \rightarrow p + \bar{\nu}_e$ 4) $p + \bar{\nu}_e
\rightarrow n + e^+$, 5) $p + e^- + \bar{\nu}_e \rightarrow n$, 6) $p +
e^- \rightarrow n +\nu_e$ 7) The Hubble expansion rate, when the rates
fall below the expansion rate, $n/p$ ``freezes'' out.
\label{ratplot}
}
\end{figure}

Since nearly all of the neutrons present during primordial
nucleosynthesis go into $^4$He nuclei, the production of primordial
helium in the early universe is dominated by the abundance of neutrons
relative to protons.  The weak nucleon interconversion reactions 
$$ 
n + \nu_e \rightleftharpoons p + e^- \;\;\;\;\; n + e^+ \rightleftharpoons p +
\bar{\nu}_e \;\;\;\;\; n \rightleftharpoons p + e^- + \bar{\nu}_e .  
$$
are in steady state equilibrium at temperatures $T > 0.7 \,\rm MeV$.
The neutron-to-proton ratio $(n/p)$ freezes out of equilibrium when
the weak nucleon interconversion rates fall below the expansion rate
(see Fig. \ref{ratplot}).  The ``frozen'' $n/p$ ratio slowly evolves
as free neutrons decay until the epoch ($T\sim 0.1$ MeV) when almost
all neutrons are incorporated into alpha particles
(``nucleosynthesis'').  In steady state equilibrium, $n/p$ can be
approximated by the ratio of the nucleon interconversion rates, $$ {n
\over p} \approx {\lambda_{p\rightarrow n} \over \lambda_{n\rightarrow
p}}.$$ The evolution of $n/p$ is followed numerically in detail in
Kawano's \cite{kawano} update of Wagoner's code \cite{wfh}.  The
suppression of low energy electron-type neutrinos affects the rates of
the six weak nucleon reactions differently.

If sterile neutrinos are partially or completely thermalized, they
will increase the expansion rate since they will contribute to the
energy density of the primordial plasma.  The expansion rate, $H$, is
related to the total energy density, $\rho_{\rm tot}$ and the
cosmological constant, $\Lambda$ as \cite{wfh,kawano}  
$$H = \sqrt{{8 \pi\over 3}
G\left(\rho_{\rm tot} + {\Lambda\over 3}\right)}.$$ 
In general, an increase to
the energy density and expansion rate will increase $n/p$ at
freeze-out, and thus increase the primordial $^4$He yield, $Y_p$.  The
contribution to the expansion rate is not significant for
$\nu_e\rightarrow \nu_s(\bar{\nu}_e\rightarrow \bar{\nu}_s)$
conversion post-decoupling, since the energy density in electron-type
neutrinos is just transferred to steriles without further thermal
creation of neutrinos (since they have decoupled from the $e^\pm$
plasma before the onset of transformation).

\begin{figure}[ht]      % in second brace, h=here, t=top, b=bottom      
\centerline{\epsfxsize 5 truein \epsfbox{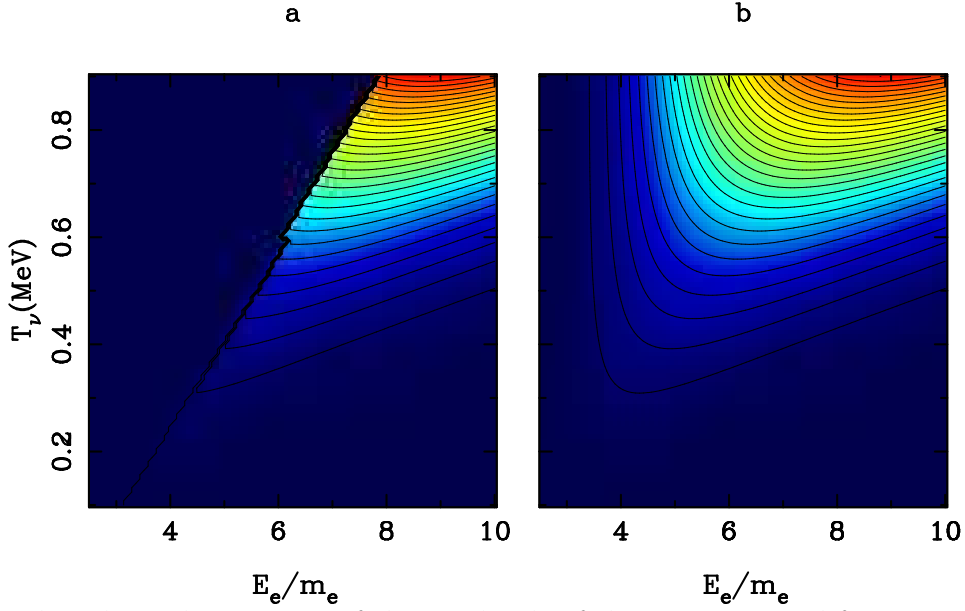}}   
\vskip 0 cm
\caption[]{
\label{intnue}
\small Plotted are the contours of the amplitude of the rate
integrand for $n + \nu_e \rightarrow p + e^-$, with increasing amplitude
towards the upper right: (a) shows the blocking of the integrand for lower
energies due to the suppression of low energy electron neutrinos;  (b) shows
the standard BBN integrand.}
\end{figure}

\begin{figure}[ht]     
\centerline{\epsfxsize 5.5 truein \epsfbox{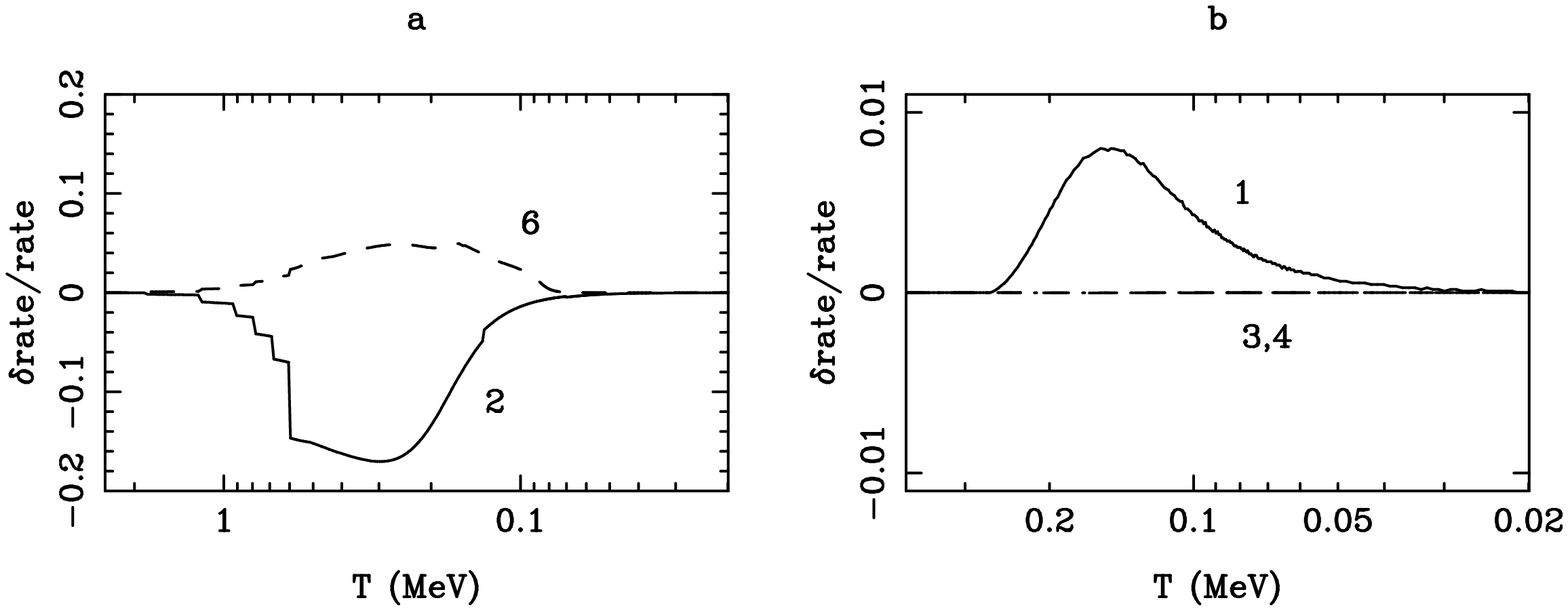}}   
\vskip 0 cm
\caption[]{
\label{deviate}
\small Plotted are a) the change in rate 2. $n + \nu_e \rightarrow p +
e^-$, 6. $p + e^- \rightarrow n +\nu_e$; and b) 1. $n \rightarrow p +
e^- + \bar{\nu}_e$, 3. $n + e^+ \rightarrow p + \bar{\nu}_e$ 4. $p +
\bar{\nu}_e \rightarrow n + e^+$ }
\end{figure}

For the case where $L_{\nu_e} < 0$, the reactions influenced by the
$\nu_e$ distortion are the forward and reverse reactions of $n + \nu_e
\rightleftharpoons p + e^-$.  Because of the neutron/proton mass
difference, very low energy neutrinos readily participate in $n +
\nu_e \rightarrow p + e^-$. Here, the neutrino energy is 
$E_\nu = E_e - Q > 0$, where $Q \equiv m_n - m_p$ is the nucleon mass
difference, and $E_e$ is the electron energy.  The suppression of low
energy neutrinos impacts this reaction by significantly limiting the
phase space of initial states of the interaction.  The neutrino energy
distribution is accounted for in the  $n + \nu_e
\rightarrow p + e^-$ rate integral in the following manner:
\begin{equation}
\label{rate2}
\lambda_{n \nu \rightarrow p e} = A \int{v_e E_e^2 p_\nu^2 dp_\nu
[e^{E_\nu/k T_\nu} + 1]^{-1} [1+e^{-E_e/k T}]^{-1}}.
\end {equation}
The rate of the reaction $p + e^- \rightarrow n + \nu_e$ is given by
\begin{equation}
\label{rate0}
\lambda_{pe\rightarrow n\nu} = A \int{E_\nu^2 p_e^2 dp_e
[e^{E_e/kT} +1]^{-1} [1+e^{-E_\nu/k T_\nu}]^{-1}}.
\end{equation}
In these equations we have used Weinberg's notation \cite{weinberg}.
The deficit in low energy neutrinos renders the integrand of
(\ref{rate2}) to be nearly zero for energy ranges where $\nu_e$ have
transformed to steriles (see Fig. \ref{intnue}). The reaction $p + e^-
\rightarrow n + \nu_e$ is altered by a lifting of Fermi-blocking
$[1+e^{-E_\nu/kT_\nu}]^{-1}$ at low energies because of the reduction of
low energy neutrino numbers.  However, the rate $\lambda_{pe\rightarrow n
\nu}$ is changed less significantly than $\lambda_{n \nu \rightarrow p
e}$ and thus affects $n/p$ less.  This can be seen in
Fig. \ref{deviate} (a), where the change in rate with respect to the
overall $n\rightarrow p (p\rightarrow n)$ rate is shown for rate (2)
$\lambda_{n \nu \rightarrow p e}$ and (6) $\lambda_{p e\rightarrow n
\nu}$.

\begin{figure}[ht]     
\centerline{\epsfxsize 5 truein \epsfbox{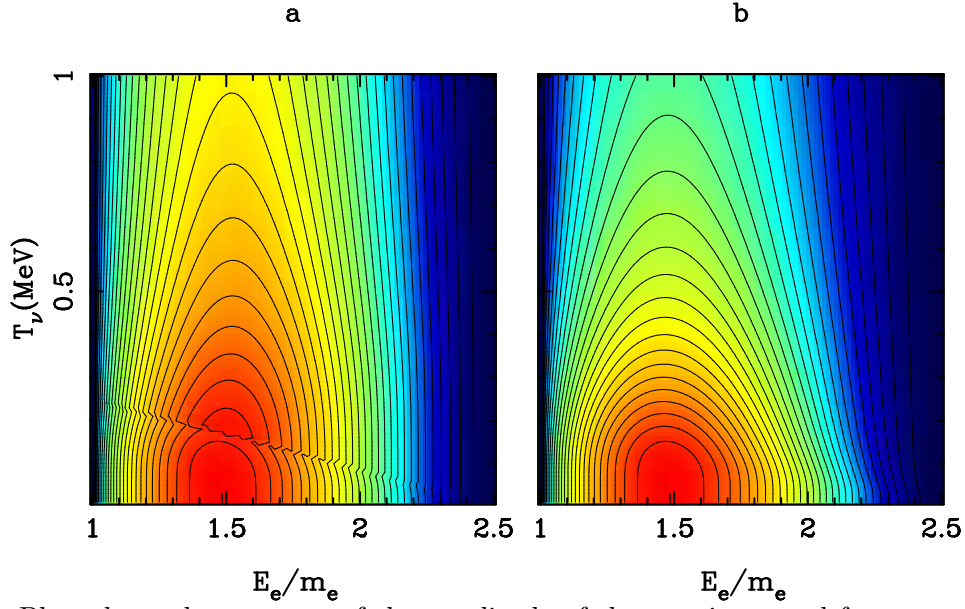}}   
\vskip 0 cm
\caption[]{
\label{intnueb}
\small 
Plotted are the contours of the amplitude of the rate
integrand for $n \rightarrow p + e^- + \bar{\nu}_e$, with increasing
amplitude towards the center. (a) shows the lack of Fermi blocking by
low energy $\bar{\nu}_e$ at higher temperatures.  (b) shows
the standard BBN integrand.}
\end{figure}

For $L_{\nu_e} > 0$, the reactions affected by $\bar{\nu}_e$
distortion are $n + e^+ \rightleftharpoons p + \bar{\nu}_e$ and $n
\rightleftharpoons p + e^- + \bar{\nu}_e$.  In the forward and reverse
reactions $n + e^+ \rightleftharpoons p + \bar{\nu}_e$, the escaping
or incident $\bar{\nu}_e$ must have an energy $E_{\bar{\nu}_e} = Q +
E_e \ge 1.804 \, \rm MeV$.  Even for the case we examined with the
greatest spectral distortion  ($|\delta m_{se}^2| = 1\,\rm eV^2$), the
spectral cutoff never extends above $E_{\bar{\nu}_e} = 1.35\,\rm
MeV$. Therefore, $n + e^+ \rightleftharpoons p + \bar{\nu}_e$ is not
altered by this low energy distortion (see Fig. \ref{deviate} (b)).
The interaction $p + e^- + \bar{\nu}_e \rightarrow n$ makes a very
small contribution to $\lambda_{p\rightarrow n}$, and modifications of
its rate are not important to $n/p$ at freeze-out.  The only
significant reaction rate affected by the $\bar{\nu}_e$ distortion is
neutron decay: $n \rightarrow p + e^- + \bar{\nu}_e$
(Fig. \ref{deviate}).  The reaction is limited in SBBN by
Fermi-blocking of low-energy $\bar{\nu}_e$ products.  This blocking is
lifted when low-energy $\bar{\nu}_e$ transform to $\nu_s$.  The
integrand of the neutron decay rate
\begin{equation}
\label{rate1}
\lambda_{n\rightarrow p e \nu} = A \int{v_e E_\nu^2 E_e^2 dp_\nu
[1+e^{-E_\nu/k T_\nu}]^{-1} [1+e^{-E_e/k T_\nu}]^{-1}}
\end {equation}
can be seen in Fig. \ref{intnueb} for the most dramatic case of
$|\delta m^2| = 1\,\rm eV^2$.  The Fermi-blocking term, $[1+e^{-E_\nu/k
T_\nu}]^{-1}$, is unity for low energy neutrinos that have
transformed to steriles.

The change in $\lambda_{n \nu \rightarrow p e}$ (for $L_{\nu_e} < 0$)
is significantly larger than that for $\lambda_{n\rightarrow p e \nu}$
(for $L_{\nu_e} > 0$).  The change in helium mass fraction ($Y_p$)
produced by BBN for $L_{\nu_e} < 0$ is $\delta Y_p \simeq 0.022$,
while a $L_{\nu_e} > 0$ correspondingly causes smaller a change of
$\delta Y_p \simeq -0.0021$ \cite{xfalept}.

In summary, if electron-type neutrino spectra are altered through
matter-enhanced transformation and survive until nucleosynthesis, they
can modify $n\rightleftharpoons p$ interconversion rates through the
availability of initial and final neutrino energy states.  This kind
of scenario is realized in $\nu_e\rightarrow
\nu_s(\bar{\nu}_e\rightarrow \bar{\nu}_s)$ matter-enhanced
transformations with $|\delta m^2| \le 1\rm\, eV^2$.  Non-thermal
$\nu_e/\bar{\nu}_e$ spectra will arise in any case where flavor
transformations occur near or below $T_{\rm dec}$ between a
more-populated electron-type neutrino and some other less-populated
neutrino flavor, and will produce the same kinds of limitations on
reaction phase space as the resonance point goes from lower to higher
neutrino energies.  The effects of non-thermal electron-type neutrino
distributions will then need to be included.

\acknowledgments

K.~A., X.~S., and G.~M.~F. are partially supported by NSF grant
PHY98-00980 at UCSD.  K.~A. acknowledges partial support from the NASA
GSRP program.

\newpage

\begin{references}

%\bibitem{nnu1} F.~Hoyle and R.~J.~Tayler, Nature {\bf 203}, 1108
%(1964); P.~J.~E.~Peebles, Phys. Rev. Lett. {\bf 16}, 411 (1966);
%V.~F.~Shvartsman, Pis'ma Zh. \'Eksp. Teor. Fiz. {\bf 9}, 315 (1969)
%[JETP Lett. {\bf 9}, 184 (1969)].

\bibitem{nnu2} G.~Steigman, D.~N.~Schramm \& J.~R.~Gunn, {\it
Phys. Lett.} {\bf B 66}, 202 (1977).

\bibitem{zdecay} Particle Data Group, {\it Review of Particle Physics}, 
{\it European Physical Journal} {\bf C3}, 1 (1998).

\bibitem{caldconrad} D.~O.~Caldwell, {\it Int.~J.~Mod.~Phys.} {\bf A13},
4409 (1998); J.~M.~Conrad, Plenary Talk presented at the International
Conference on High Energy Physics (ICHEP), Vancouver, 1998,
hep-ex/9811009.

\bibitem{dolgov81} A.~D.~Dolgov, {\it Sov. J. Nucl. Phys.} {\bf 33}, 5.

\bibitem{barbdolg} R.~Barbieri and A.~Dolgov,
{\it Phys. Lett.} {\bf B 237}, 440 (1990); {\it ibid.},
{\it Nucl. Phys.} {\bf B 349}, 743 (1991). 

\bibitem{enqvist} K.~Enqvist, K.~Kainulainen and M.~Thomson,
{\it Nuc. Phys.} {\bf B 373}, 498 (1992).

\bibitem{xsf} X.~Shi, D.~N.~Schramm and B.~D.~Fields,
{\it Phys. Rev.} {\bf D48}, 2563 (1993).

\bibitem{SMF} M.~J.~Savage, R.~A.~Malaney \& G.~M.~Fuller,  \apj
{\bf 368}, 1 (1991).

\bibitem{alpher} R.~A.~Alpher, J.~W.~Follin,~Jr. and R.~C.~Herman,
{\it Phys. Rev} {\bf 92}, 1347 (1953).

\bibitem{wfh} R.V.~Wagoner, W.A.~Fowler and F.~Hoyle, 
{\it Astrophys. J.} {\bf 148} 3 (1967).

\bibitem{kirchiz} D.~P.~Kirilova \& M.~V.~Chizhov, {\it Phys. Rev.}
{\bf D58}, 073004 (1998).

\bibitem{xfalept} X.~Shi, G.~M.~Fuller and K.~Abazajian, {\it
Phys. Rev.} {\bf D 60}, 063002 (1999).

\bibitem{mckthom94} B.~McKellar, and M.~J.~Thomson, {\it Phys. Rev.} {\bf
D 49}, 2710 (1994).

\bibitem{fullmalan} G.~M.~Fuller \& R.~A.~Malaney.  {\it Phys. Rev.} {\bf D
43}, 3136 (1991).

\bibitem{foot} R.~Foot and R.~R.~Volkas, {\it Phys. Rev.} {\bf D 56}, 6653
(1997).

\bibitem{OSS} K.~Olive, G.~Steigman and E.~Skillman,
{\it Astrophys. J.} {\bf 483}, 788 (1997).

\bibitem{burtyt} S.~Burles and D.~Tytler, \apj {\bf 499}, 699
(1998); {\it ibid.}, \apj {\bf 507}, 732 (1998); D.~Tytler,
X.~Fan and S.~Burles, {\it Nature} {\bf 381}, 207 (1996).

\bibitem{xdchaos} X.~Shi, {\it Phys. Rev.} {\bf D 54}, 2753 (1996).  

\bibitem{xdcaus} X.~Shi and G.~M.~Fuller astro-ph/9904041 (unpublished).

\bibitem{kawano} L.~Kawano, FERMILAB-PUB-92-04-A, Jan 1992.

\bibitem{weinberg} S.~Weinberg, {\it Gravitation and Cosmology},
J. Wiley, New York (1972).

\end{references}
\end{document}